\documentclass[twocolumn,amssymb, amsmath, aps, showpacs, footinbib, prb]{revtex4-1}

\usepackage{graphicx}
\usepackage{dcolumn}
\usepackage{bm}
\usepackage{color}

\newcommand{\degree}{\ensuremath{^\circ}}

\begin{document}


\title{ Interplay of antiferromagnetism, ferromagnetism and superconductivity in EuFe$_2$(As$_{1-x}$P$_x$)$_2$ single crystals }

\author{H. S. Jeevan,$^1$ Deepa Kasinathan,$^2$ Helge Rosner,$^2$ and Philipp Gegenwart$^1$}

\affiliation{$^1$I. Physikalisches Institut, Georg-August-Universit\"at G\"ottingen, D-37077 G\"ottingen, Germany}
\affiliation{$^2$Max-Planck-Institut f\"ur Chemische Physik fester Stoffe, 01187 Dresden, Germany}

\begin{abstract}
We report a systematic study on the influence of
antiferromagnetic and ferromagnetic phases of Eu$^{2+}$ moments on the
superconducting phase upon doping the As site by isovalent P, which essentially acts like chemical pressure on EuFe$_2$As$_2$.
 Bulk superconductivity with transition temperatures of 22\,K and 28\,K are observed for $x$=0.16 and 0.20 samples,
 respectively. The Eu ions order antiferromagnetically for $x$\,$\leq$\,0.13, while a crossover is observed for
 $x$\,$\geq$\,0.22 whereupon the Eu ions order
 ferromagnetically. Density functional theory based calculations reproduce the observed experimental findings
 consistently. We discuss in detail the coexistence of superconductivity and magnetism in a tiny region of 
 the phase space and comment on the competition of ferromagnetism and superconductivity in the title compound.

\end{abstract}


\maketitle

The appearance of superconductivity (SC) in the vicinity of a magnetic
instability is often related to quantum critical phenomena,\cite{mathur,stockert}
although only in few cases the magnetic excitations which mediate the SC 
pairing have been identified.\cite{sato}
The discovery
of superconductivity upon suppression of magnetism in iron containing pnictides and
chalcogenides has created great interest
in the field of condensed matter physics. Among the various members of
the iron containing pnictides, there are
three main family of materials, which show SC
transitions upon substitution by a dopant or upon applying external pressure.
They are, (i) the quaternary `1111' compounds, {\it R}FeAsO, where
{\it R} represents a lanthanide such as La, Ce, Sm etc. \cite{Kamihara,Takahashi,Chen,chenCe} with 
transition temperatures as high as 56\,K; (ii) the ternary
{\it A}Fe$_2$As$_2$ ($A$ = Ca, Sr, Ba, Eu) \cite{RotterBa,sasmalSr,Jeevan2,WuCa} systems, also known as
`122' systems that exhibit superconductivity up to 38\,K; and (iii) the binary chalcogenide `11' systems ($eg$. FeSe)
with superconducting transition temperatures upto 14\,K.\cite{FeSe}
In general, the magnetism occurring in the Fe sublattice can be suppressed by 
doping via two schemes: (i) direct doping of Fe in the FeAs layer by Co, Ni, Rh, (electron doping)\cite{alj,deepa} or Ru (isovalent substitution)\cite{deepa2}
(ii) indirect doping on other sites which includes, oxygen by fluorine in the `1111' systems (electron doping),\cite{Kamihara} alkaline earth metals by alkaline metals in the `122' systems (hole doping),\cite{RotterBa} and arsenic with phosphorus 
(isovalent substitution).\cite{Cao1,Cao2}
Similar to doping, external pressure also facilitates the suppression of Fe magnetism.\cite{Miclea} In the case of 
rare-earth based iron pnictides, a second magnetic sublattice, due to the localized {\it f}-moments comes into play
additionally to the Fe sublattice. In general, the rare-earth ions tend to order antiferromagnetically, thereby 
introducing only a weak coupling between the two sublattices.  
In this work we concentrate on EuFe$_2$As$_2$, the only rare-earth based member of the `122' family.
EuFe$_{2}$As$_{2}$ exhibits a spin density
wave (SDW) in the Fe sublattice together with a structural  transition at 190\,K and in addition an
A-type antiferromagnetic (AF) order at 19\,K due to Eu$^{2+}$ ions (ferromagnetic 
layers ordered antiferromagnetically).\cite{Jeevan1}
Superconductivity can be achieved in this system by
substituting Eu with K or Na (Ref.\,\onlinecite{Jeevan2,QiNa}),
As with P (Ref.\,\onlinecite{RenP}) and upon application of external pressure.\cite{Miclea,AFMSC}
Pressure studies upto 3 GPa on the parent compound have also shown indications of {\it reentrant} SC, akin
to ternary Chevrel phases or rare-earth nickel borocarbides.\cite{Miclea}

Isovalent P doping on the As site in EuFe$_2$As$_2$ without introducing
holes or electrons, simulates a scenario generally
referred to as ``chemical pressure".
While the Eu$^{2+}$ moments order antiferromagnetically (A-type)
at 19\,K in the parent compound, ferromagnetic order at 27\,K is found for
the end member EuFe$_{2}$P$_{2}$. \cite{EuFeP}
In early 2009, Ren and coworkers\cite{RenP} reported on the co-existence of SC and
ferromagnetism (FM) of the Eu$^{2+}$ moments in polycrystalline samples of
EuFe$_2$(As$_{0.7}$P$_{0.3}$)$_{2}$, with a superconducting transition at 26\,K, followed
by a FM ordering of the Eu$^{2+}$ moments at 20\,K. 
Recently, another report\cite{ahmed} also documents the co-existence of  SC ( at 26\,K) and FM (at 18\,K) 
in EuFe$_2$(As$_{0.73}$P$_{0.27}$)$_{2}$ along with a reentrant behavior below 16\,K.
On the other hand systematic
studies by Ren and collaborators on Ni doping in EuFe$_{2-x}$Ni$_{x}$As$_2$ showed only
FM ordering of the Eu$^{2+}$ moments but no superconductivity.\cite{NidopeEu}
In contrast, superconductivity has been reported upon Ni doping of the Fe site
for the other three members of the $A$Fe$_{2-x}$Ni$_x$As$_2$ ($A$ = Ca, Sr, Ba)
family.\cite{saha,li,neeraj,deepa}
Based on these reports, the physical properties of both Ni
and P doped EuFe$_2$As$_2$ samples seem to contradict each other in terms of competition or
coexistence of FM and SC phases. In this paper, we
report on the detailed investigation of the
resistivity, magnetization and specific heat measurements using
well-characterized single crystals of EuFe$_2$(As$_{1-x}$P$_x$)$_2$ and show the presence
of bulk superconductivity up to
28\,K.
Our measurements also prove that in this
system, the FM (Eu$^{2+}$ ions) and SC phases compete with each
other, rather than coexist, in contradiction to the claim of Ref.~\onlinecite{RenP,ahmed}.
Bulk SC coexisting with AF Eu$^{2+}$ ordering is only found in a very narrow regime of 
P doping, where the Fe SDW transition has just been suppressed. 
The systematic study of P doping on single crystals allows us to draw a phase
diagram that is complex and rich with five different phases.

A series of single crystals of EuFe$_2$(As$_{1-x}$P$_x$)$_2$ with
a range of P doping were synthesized using the
Bridgman method. Stoichiometric amounts of starting elements (Eu
99.99$\%$, Fe 99.99 $\%$, As 99.99999$\%$ and P 99.99 $\%$) were
taken in an Al$_{2}$O$_{3}$ crucible, which was then sealed in a
Ta-crucible under Argon atmosphere. The sealed crucible was heated
 at a rate of 50$\degree$C/hour up to 1300$\degree$C, kept
for 12 hours at the same temperature and then cooled to 950$\degree$C with a cooling
rate of 3$\degree$C/hour. We obtained large plate-like single crystals
using this process with dimensions of 5 x 3 mm$^2$  in the
$ab$-plane. In addition to the plate-like single crystals, we also found a
secondary polycrystalline phase which was identified as Fe$_{2}$P. 
All the elements and sample handling were
carried out inside a  glove box filled with Ar atmosphere. The quality of the single crystals was
checked using the Laue method,
powder x-ray diffraction and additionally with scanning electron microscopy
equipped with energy dispersive x-ray analysis (EDX). Electrical
resistivity and specific heat were measured using a
Physical Properties Measurement System (PPMS, Quantum Design, USA).
Magnetic properties were measured using Superconducting Quantum
Interference Device (SQUID) magnetometer procured from Quantum
Design. The change of the electronic properties with phosphorus doping
has been investigated by angle-resolved photoemission spectroscopy (ARPES) and 
optical conductivity measurements\cite{fink1,fink2,wu} on our single crystals. 
ARPES measurements indicate that the electronic structure of the phosphorus
doped systems are more 3D compared to the parent compound and as well
as to the case of electron doping.\cite{fink1,fink2}
Recent infrared spectroscopy measurements claim evidence for a single nodeless s-wave 
superconducting gap for isovalent substitution in contrast to the multi-gap scenario
in carrier doped systems.\cite{wu} 
In this paper we focus on the 
interplay of Eu$^{2+}$ magnetism on the formation of SC.

\begin{table}[t]
\caption{\label{lattice} Tetragonal lattice parameters (for selected samples at 300\,K) of EuFe$_{2}$(As$_{1-x}$P$_{x}$)$_{2}$ 
crystallizing in the ThCr$_{2}$Si$_{2}$-type structure as a function of phosphorus substitution ($x$), which 
is determined by EDX }
\begin{ruledtabular}
\begin{tabular}{@{\extracolsep{\fill}}lll}
$x$     &  $a$\,(\AA) & $c$\,(\AA) \\
\hline
0            &   3.907   & 12.114 \\
0.15      &   3.891   & 11.948\\
0.16      &   3.890     & 11.930\\
0.20      &  3.887   & 11.890\\
0.22     & 3.889   & 11.877\\
0.26       & 3.889   & 11.870\\
0.38	     & 3.885    &11.774\\
1            & 3.816   & 11.248\\
\end{tabular}
\end{ruledtabular}
\end{table}

The powder diffraction pattern of all the
samples can be indexed using the tetragonal ThCr$_2$Si$_2$ structure type. The variation of the
 lattice parameters with respect to P content is collected in Table.\,I,
wherein the actual P content is given as determined by EDX. 
Similar to other isovalent substitutions of 122 systems discussed in literature\cite{BaFeAsP}, phosphorus
substitution on the As site leads to a dramatic decrease of FeAs-layer thickness, which
indicates that P doping  mainly affects the $c$ lattice parameter. 
We also observe that the decrease along the tetragonal axis ($c$ lattice parameter) is more pronounced
compared to the changes in the $ab$ plane ({\it cf.} Table. I).

\begin{figure}[t]
\begin{center}
\includegraphics[angle=-90,width= 8.5cm,clip]{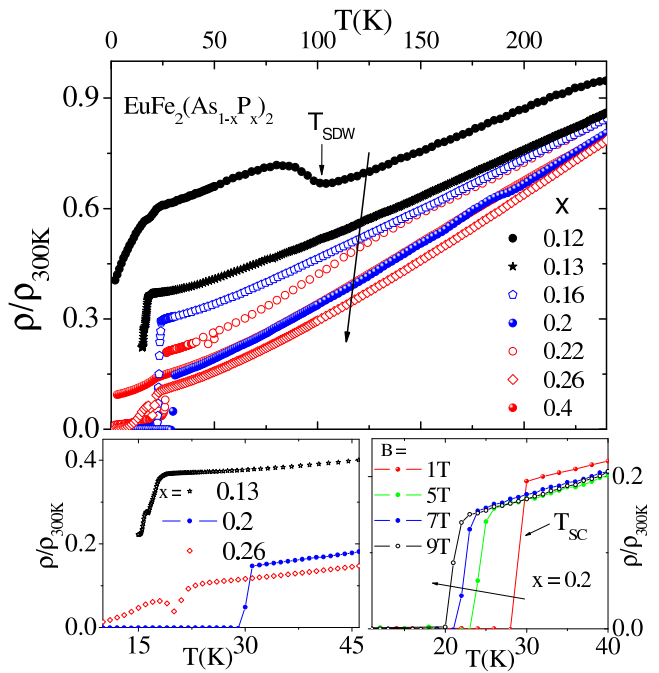}
\caption{\label{resistivity}(Color online) Temperature dependence of the 
in-plane ($ab$-plane) resistivity for various EuFe$_2$(As$_{1-x}$P$_x$)$_2$ single crystals.
The data are normalized to the room temperature resistivity. Due to technical reasons, 
the resistivity for $x$=0.13 sample was measured down to 15\,K only. The lack of SC (zero resistivity) 
transition is confirmed by the magnetic susceptibility measurements (Fig.~\ref{susceptibility}). Lower left panel:  Low temperature part
of the normalized in-plane resistivity for selected samples. Lower right panel: Normalized in-plane resistivity
of the superconducting sample ($x$=0.2) for various values of applied magnetic field.}
\end{center}
\end{figure}

The temperature dependence of the normalized resistivity (with the current measured in the basal $ab$-plane) is shown in
Fig.~\ref{resistivity} for the single crystals of EuFe$_2$(As$_{1-x}$P$_x$)$_2$.
The temperature dependence of the resistivity shows a metallic behavior for the entire doping range.
For the parent compound EuFe$_2$As$_2$, we observe both the SDW anomaly associated with the 
structural and magnetic transition of the Fe sublattice\cite{Jeevan1} at 
190\,K (T$_{SDW}$) and the anomaly at 19\,K (T$_{N}$) associated with the AF ordering of the Eu$^{2+}$ moments.
Upon doping with P,  for the lowest P content studied here
$x$=0.12, the resistivity decreases linearly with
temperature down to 90\,K, whereupon we observe an anomaly which is likely due to the SDW transition.  
We observe another anomaly around 20\,K which is likely associated with the A-type AF ordering of
the Eu$^{2+}$ ions. 
No further 
anomalies are observed for this sample.
When the P content is increased, the SDW transition is fully suppressed and a sudden drop in the resistivity indicative of a superconducting (SC) transition
is observed for $x$=0.16 at 22\,K and for $x$=0.2 at 29\,K.
For $x$=0.22, a sharp drop in the resistivity is observed around 25\,K but the normalized resistivity does not go to zero. 
For larger P concentrations, $x$\,$\geq$\,0.26, the SC transition is fully suppressed. 
Our observations described here are in contradiction to previous reports\cite{RenP,ahmed} which
evidence a SC transition around 26\,K for polycrystalline samples with $x$=0.27 and 0.30. We do not observe
zero resistance consistently for all samples with $x$\,$\geq$\,0.22.
The lower right panel of Fig.~\ref{resistivity} shows the field dependence of the normalized resistivity of the SC transition temperature for the
$x$=0.2 sample. The $\rho$(T)/$\rho_{300K}$
shows a sharp drop below 30\,K in zero magnetic field and shifts to lower temperatures
as the field increases and reaches 22\,K at 9\,T, in accordance with the SC properties.

\begin{figure}[t]
\begin{center}
\includegraphics[width= 8.5cm]{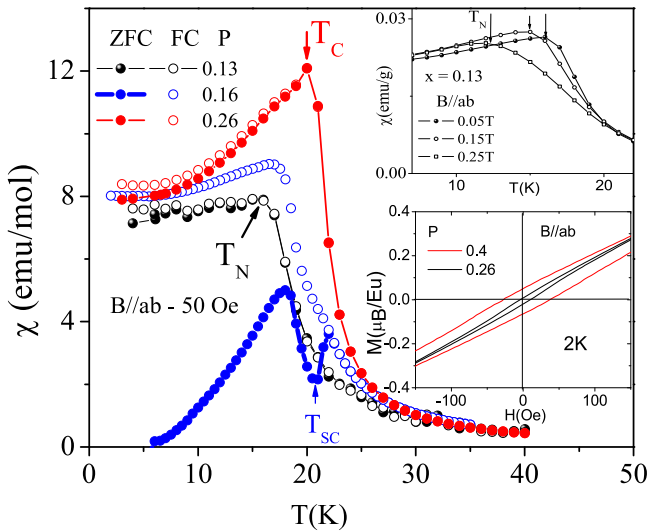}
\caption{\label{susceptibility} (Color online) Temperature dependence of the magnetic
susceptibility for EuFe$_2$(As$_{1-x}$P$_x$)$_2$.  Upper inset: Field dependence of the AF transition for the $x$=0.13 sample.
Lower inset:  Isothermal magnetization at 2\,K of $x$=0.26 and 0.4 samples which show a FM ordering of the Eu$^{2+}$ moments.}
\end{center}
\end{figure}

We measured the dc magnetic susceptibility for selected compositions to obtain more insights
regarding the nature of the magnetic and SC phases, with the applied magnetic field parallel to the $\it{ab}$ plane.
Fig.~\ref{susceptibility} shows the field cooled (FC) and zero field cooled (ZFC)
susceptibility at 50\,Oe, for $x$ = 0.13, 0.16 and 0.26. 
For the under-doped sample ($x$=0.13) both ZFC and FC data show only the anomaly at 17\,K due to AF ordering of
the Eu$^{2+}$ moments.
There are no signs for a SC phase for this sample.
For the optimally doped $x$=0.16 sample, the susceptibility shows a pronounced diamagnetic step at 22\,K, evidence for a bulk SC transition, consistent with
the sharp drop observed in resistivity below 22\,K. Above 20\,K the susceptibility begins to increase and reaches a 
maximum around 18\,K, which is indicative of the AF ordering of the Eu$^{2+}$ moments. This observation of the 
coexistence of SC and AF ordering for $x$=0.16 is similar to the ac-susceptibility measurements reported previously for 
the parent compound under 25.7 kbar pressure.\cite{AFMSC} 
Further increasing the P content to $x$=0.26, the magnetic susceptibility shows only an anomaly at 20\,K due to
the FM ordering of the Eu$^{2+}$ moments (discussed later). 
It is clear from the magnetic susceptibility measurements, that bulk SC phase transition
is observed
only for the $x$=0.16 sample, while no sign for a SC phase is inferred for $x$= 0.13 and 0.26 samples.
The inset of Fig.~\ref{susceptibility} shows the field dependence of T$_N$ for the under-doped sample, $x$=0.13. With increasing fields, T$_{N}$
is shifted to lower values. This is indicative of the AF ordering of the Eu$^{2+}$ moments.
Upon increasing the P content to $x$=0.26, the Eu ordering changes from AF to FM.
The FM ordering of the $x$=0.26 and 0.4 samples are confirmed by isothermal magnetization
measurements at 2\,K shown in the lower inset of Fig.~\ref{susceptibility}.
A small but clear hysteresis loop is observed as a function of applied field, which is consistent
with similar observation in the ferromagnetically (T$_C$=29\,K) ordered end member EuFe$_2$P$_2$.\cite{EuFeP}
Previous reports\cite{RenP,ahmed} have claimed the coexistence of SC with FM for $x$=0.27 and 0.3 polycrystals,
which display a tiny (10$^{-2}$\,emu/mol) diamagnetic contribution to the susceptibility together with a
broadened resistive transition at ~25\,K and which also shows a re-entrance behavior around 16\,K.
The data on single crystals presented here indicate clear bulk SC only within the concentration range 0.16\,$\leq$\,$x$\,$\leq$\,0.2.
At large P doping, where Eu displays FM ordering, SC appears to be suppressed.
\begin{figure}[t]
\begin{center}
\includegraphics[width= 8.5cm]{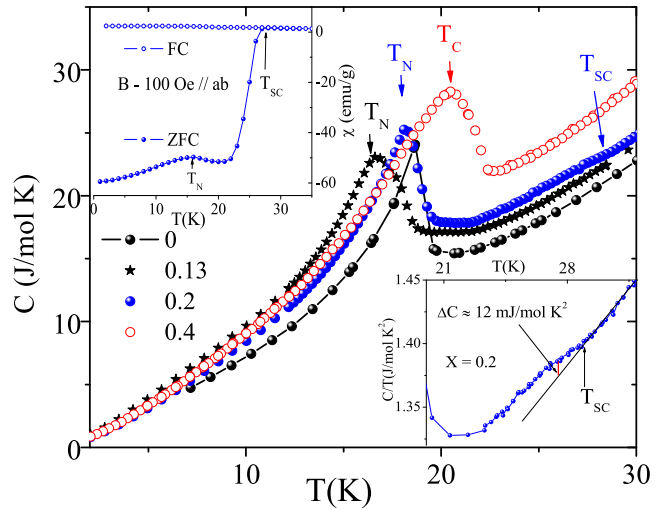}
\caption{\label{specificheat} (Color online) Evaluation of AF (T$_{N}$) and FM (T$_{C}$) ordering temperatures as a function of 
phosphorus substitution in the specific heat. Lower inset shows the anomaly of the SC phase transition for the
$x$=0.2 sample. Upper inset shows the dc-magnetic susceptibility for ZFC and FC experiments in applied field of 50 Oe for the
$x$=0.2 sample. }
\end{center}
\end{figure}

Another confirmation for the bulk nature of the magnetic and SC phases is obtained by measuring the specific heat (C$_{p}$)
in the temperature range from 35\,K down to 2\,K. The data are collected in Fig.~\ref{specificheat} for an
under-doped ($x$=0.13), optimally doped ($x$=0.2) and over-doped ($x$=0.4) single crystal.
The plot shows clear anomalies for both AFM (T$_{N}$) and FM (T$_{C}$) phase transitions for $x$=0.13 and 0.4 respectively. 
Due to strong contributions at low temperatures from the phonons and Eu$^{2+}$ magnetic moments to the specific heat,
it is difficult to observe an anomaly at the SC phase transition, however a small anomaly
(lower inset of Fig.~\ref{specificheat}) in the raw data (without subtracting any phonon
contribution) is resolved below 28\,K for $x$=0.2. This observation is consistent with the diamagnetic step observed in the magnetic
susceptibility (upper inset of Fig.~\ref{specificheat}) as well as the drop in resistivity (Fig.~\ref{resistivity}). 
In addition, the magnetic susceptibility for $x$=0.2 (ZFC at 50\,Oe) shows a small anomaly at
20\,K corresponding to the AF transition of the Eu$^{2+}$ moments, indicative of a co-existence of AF and SC for this
sample. 
Based on these thermodynamic measurements, we infer that both the magnetism and SC are of bulk nature and furthermore 
only AF and SC phases co-exist for  EuFe$_2$(As$_{1-x}$P$_x$)$_2$.
A single crystal with $x$=0.18, displaying a SC transition at 28\,K has also been studied by optical conductivity measurements\cite{wu} 
above and below T$_{C}$. A BCS fit revealed a s-wave type gap with 2$\Delta$\,=\,3.8k$_{B}$T$_{C}$.

\begin{figure}[t]
\begin{center}
\includegraphics[width= 8.5cm]{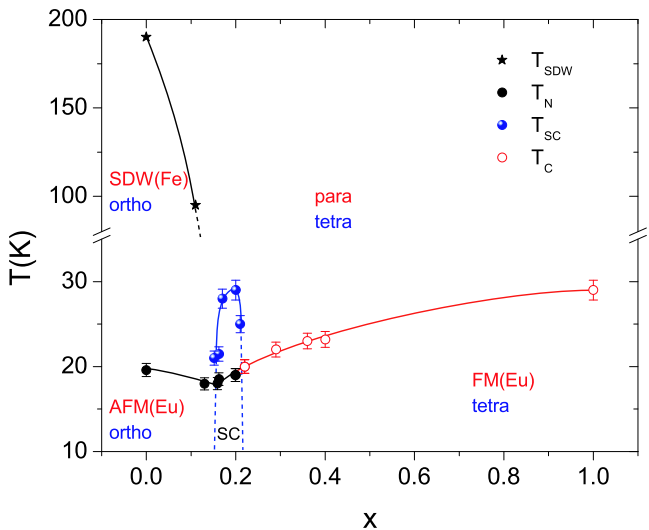}
\caption{\label{phasediag}(Color online) The complex phase diagram for
EuFe$_2$(As$_{1-x}$P$_x$)$_2$ as a function of P. The solid lines act as a guide to the eye and the dotted lines are extrapolation
using the available experimental data to close the superconducting dome.}
\end{center}
\end{figure}

The results obtained from the different measurements allow us to draw the
electronic phase diagram of  EuFe$_2$(As$_{1-x}$P$_x$)$_2$ as a function of phosphorus doping. 
The transition temperatures were determined from  specific heat, resistivity and
magnetic susceptibility measurements. Fig.~\ref{phasediag} shows a complex phase diagram, wherein we have identified five
different phases. The parent compound EuFe$_2$As$_2$ with tetragonal symmetry is paramagnetic at high
temperatures (above 190\,K), while below 190\,K, the structure changes to orthorhombic and the Fe moments
order antiferromagnetically (SDW). In addition, below 19\,K, the system undergoes another AF transition associated
with the Eu$^{2+}$ moments.
Upon isovalent doping of the As site with P in EuFe$_{2}$(As$_{1-x}$P$_{x}$)$_{2}$, for 0.16\,$\leq$\,$x$\,$\leq$ 0.22 
we observe bulk SC phase transitions coexisting with AF ordering of the Eu$^{2+}$ moments. Further increasing the P content, $x\geq$0.26,
SC is completely suppressed and the Eu$^{2+}$ sublattice magnetic interaction changes from AFM to FM.
Recently, Nandi {\it et. al.} proposed\cite{nandi} that the coupling between
orthorhombicity and superconductivity is indirect and claim that it arises due to
the strong competition between magnetism (of Fe) and superconductivity in Co
doped BaFe$_2$As$_2$ systems. This means that the orthorhombic to
tetragonal transition occurs at temperatures above the onset of
Fe magnetic (SDW) order and the orthorhombic structure could continue to
exist in the superconducting phase too. Similar arguments can be used
for the current phase diagram of EuFe$_2$(As$_{1-x}$P$_{x}$)$_{2}$ where most likely AF
(Eu) and SC phase coexist with the orthorhombic phase.
Detailed analysis of the tetragonal to orthorhombic distortion on the stability of the FM and AF 
Eu$^{2+}$ sublattice on our samples is in progress.
In general, a FM phase is not favorable for SC
within s- wave pairing mechanisms, because the Zeeman effect arising due to ferromagnetism will
strongly disfavor the singlet formation,
which will eventually lead to the breakdown of the Cooper pairs.
FM ordering of Eu$^{2+}$ ions will result in an internal magnetic field of reasonable 
strength due to its large spin value S=7/2, which is detrimental to the SC occurring in the FeAs 
layers. 
Hence we believe that the previous reports\cite{RenP,ahmed} on the coexistence of SC and Eu-FM 
in EuFe$_{2}$(As$_{0.73}$P$_{0.27}$)$_{2}$ and EuFe$_{2}$(As$_{0.7}$P$_{0.3}$)$_{2}$ 
might probably be related to inhomogeneous phosphorus doping concentration in polycrystalline
samples.\cite{foot}
In general, single crystals are
compositionally more homogeneous than polycrystalline samples, and the presence of even a few percent volume fraction
of the superconducting phase in a given sample, should result in a strong drop (not zero) in the resistivity measurements. In our
single crystals, with $x$=0.26 and 0.38, we did not observe any such drop in resistivity. 
At this juncture, we would also like to note that, by crushing a single crystal of EuFe$_{2}$(As$_{0.62}$P$_{0.38}$)$_{2}$ into a powder we repeated all our measurements and did not notice any change in our descriptions ($i.e.$ only FM ordering of the
Eu$^{2+}$ moments was observed, but no SC phase transition).

\begin{figure}[t]
\begin{center}
\includegraphics[width= 8.5cm]{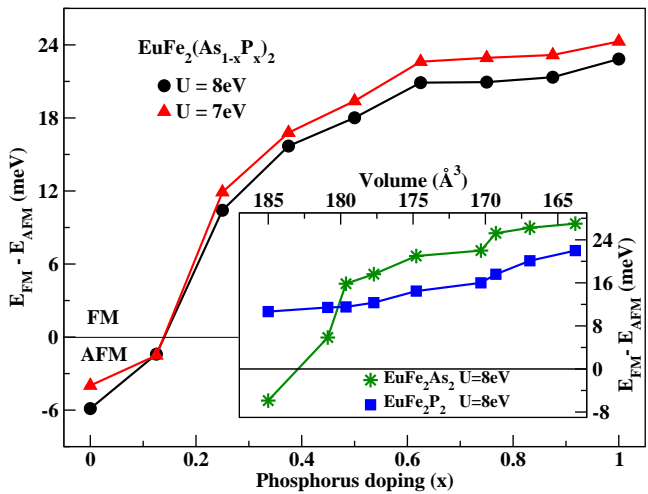}
\caption{\label{calc}(Color online) Energy difference between FM and AF aligned Eu$^{2+}$ moments along the $c$-axis,
as a function of Phosphorus content in EuFe$_2$(As$_{1-x}$P$_x$)$_2$. Along the $y$-axis, the ground state is AFM below
zero and FM above zero. Inset: Energy difference (same as main panel) as a function of reduced volume for the two end-members
EuFe$_{2}$As$_{2}$ and EuFe$_{2}$P$_{2}$. The data points in the inset and the main panel have a one-to-one correspondence.}
\end{center}
\end{figure}

For a microscopic understanding of the inter-layer coupling of the Eu$^{2+}$ moments in
EuFe$_2$(As$_{1-x}$P$_{x}$)$_{2}$, we have
carried out total energy calculations using the full-potential local orbital code (FPLO).\cite{fplo}
We used the Perdew-Wang\cite{perdew} flavor of the exchange-correlation potential and the energies were
converged  on a dense $k$-mesh consisting of 20$^{3}$ points. 
The localized Eu 4$f$ states were treated on a mean-field level by
using the LSDA+$U$ (local-spin-density-approximation + strong correlations) approach, applying the so called ``atomic limit" (AL) double-counting scheme.\cite{lsdau}
In general, the physically relevant value of the strong Coulomb repulsion $U_{\rm 4f}$ of the Eu$^{2+}$ ion are inferred 
from various spectroscopy techniques, especially by photoemission spectroscopy (PES) experiments.
Owing to the lack of such experiments for EuFe$_{2}$As$_{2}$, we have therefore used a 	 
$U_{\rm 4f}$ value
of 7-8 eV.\cite{footnote}  The robustness of our results and consequently the interpretations were checked for 
consistency with varying $U_{\rm 4f}$ values. The Fe 3$d$ states were treated on an itinerant level (LSDA)
without additional correlations.
The partial P substitution was modeled by the construction of supercells of various sizes. The randomness in possible
substitution positions were taken into account by allowing phosphorus to occupy different possible combinations 
of the four-fold 4$e$ As sites. The lattice parameters for the various supercells were obtained by interpolating linearly the 
experimental data reported in Table I. The As/P $z$-position was kept fixed at $z$=0.362 throughout.
Based on density functional theory calculations, we have previously\cite{Jeevan1} shown that the Eu and Fe
sublattices are quite de-coupled in EuFe$_{2}$As$_{2}$. This result is also corroborated experimentally
by showing that the Eu$^{2+}$ moments only play a minor role in the electronic transport properties.\cite{terashima}
Presently, our goal is to obtain an estimate for the Eu inter-layer coupling below T$_{N}$ or T$_{C}$,
while the Fe sublattice is retained in the SDW pattern.
The results from our calculations are collected in Fig.~\ref{calc}.
For $x$\,$<$\,0.2, the ground state of EuFe$_{2}$(As$_{1-x}$P$_{x}$)$_{2}$ is A-type AF, consistent with the above
mentioned experimental results. The energy difference between the AF (ground state) and FM alignment of the Eu$^{2+}$ moments
is quite small (0 to 6 meV per formula unit).  This implies a rather weak inter-layer coupling for the Eu sublattice.
Any small external effects (impurities, doping, external pressure and fields) can easily flip the Eu spins from AF to FM.
This has also been shown experimentally by Xiao and co-workers\cite{xiao} for the parent compound EuFe$_{2}$As$_{2}$, 
wherein below T$_{N}$ they observe a field-induced spin reorientations to the FM state for an applied field of just 1\,T in the
$ab$-plane and at 2\,T along the $c$-axis. 
For $x$\,$>$\,0.2, FM inter-layer coupling between the Eu$^{2+}$  moments
becomes favorable ($i.e.$ ground state is FM) and continues to remain so for larger P substitutions, consistent with the present experimental observations. The Fe sublattice remains magnetic for 0\,$\leq$\,$x$\,$\leq$\,0.875 with a slight 
reduction of the individual magnetic moments with increasing phosphorus content. The Fe sublattice becomes non-magnetic
for the end member of this substitution series EuFe$_{2}$P$_{2}$, while the rare-earth Eu remains divalent in the entire
substitution range. These results are also consistent with the recently reported experimental findings of Feng and 
co-workers\cite{EuFeP} for EuFe$_{2}$P$_{2}$.  Another recent report by Sun and collaborators\cite{sun} 
witness a valence change of europium from a 2+ to a 3+ state in EuFe$_{2}$As$_{1.4}$P$_{0.6}$ at ambient 
conditions along with a SC transition at 19.4\,K. They also suggest that the FeAs layers receive the additional charges 
arising from this valence transition, which
in turn steers the onset of SC. This valence change behavior seems rather counterintuitive, since the end member EuFe$_{2}$P$_{2}$ 
is a well known ferromagnet with divalent europium.\cite{morsen} One should also note that other alkaline-earth based members of the `122' family have been shown to superconduct
upon isovalent doping either on the Fe site or the As site, without the possibility of additional charges entering the FeAs layers.
\cite{saha,li,neeraj,deepa,deepa2}
LDA+$U$ calculations favor integer occupation, but a qualitative description of  the valence transitions can be obtained from 
such calculations.\cite{miriam}
Therefore, we investigated the possibility
of such a valence change (Eu$^{2+}$ $\rightarrow$ Eu$^{3+}$) in our calculations and found that europium always favors the divalent state in the entire
substitution range.

As mentioned earlier, without the introduction of additional holes or electrons, isovalent doping of As by P introduces
``chemical pressure" in EuFe$_{2}$(As$_{1-x}$P$_{x}$)$_{2}$. Phosphorus ion is smaller than the arsenic ion,
and the lattice parameters shrink rather anisotropically ($c/a$ decreases significantly) with increasing phosphorus content (compare Table.\,I) and the lattice becomes more 3-dimensional. Similar anisotropic changes to the lattice parameters along with a strong
decrease of the $c/a$ ratio leading to bulk SC was also
observed for isovalent substitution of Fe by larger Ru atoms.\cite{deepa2} The substitution of As by P and as well
as the decrease in the volume of the unit cell together influence the magnetism of the Eu sublattice. In order to de-couple
these two effects; substitution and volume reduction  and henceforth obtain a deeper understanding of the chemistry and
lattice effects respectively, we performed two further calculations: (i) parent compound EuFe$_{2}$As$_{2}$ as a function of reduced
volume; and (ii) end member EuFe$_{2}$P$_{2}$ as a function of reduced volume. 
The previously described supercell calculations provide information on the effects of substitution, while the current calculations 
explain the effects of 
the lattice. Our findings are summarized in the inset of Fig.~\ref{calc}. The data points in the inset have a one-to-one 
correspondence with the doping ($x$) values in the main panel. The ground state of EuFe$_{2}$As$_{2}$ changes
from AF to FM earlier than that of EuFe$_{2}$(As$_{1-x}$P$_{x}$)$_{2}$. The As 4$p$ states are more extended than 
P 3$p$ states, which when combined with a strong decrease of the $c$-axis tends to influence the inter-layer magnetic 
interaction of the Eu$^{2+}$ ions more than phosphorus.  
On the contrary, Eu$^{2+}$ moments in EuFe$_{2}$P$_{2}$ remain FM for all volumes in our calculations. 
The reported ambient conditions volume from
experiments for EuFe$_{2}$P$_{2}$ is 163.79\,\AA$^{3}$ (Ref.~\onlinecite{EuFeP}) with FM aligned Eu ions. 
Expanding this lattice to 185\,\AA$^{3}$ (room-temperature volume of EuFe$_{2}$As$_{2}$) reduces the strength of the 
FM interaction between the inter-layer Eu ions, but does not flip the spins.  
Combining these results, we infer that the lattice effects play the major role in influencing the interplay of Eu-magnetism in 
EuFe$_{2}$(As$_{1-x}$P$_{x}$)$_{2}$.

In summary, we have systematically grown high quality single crystals and  studied the transport,
magnetic and thermodynamic  properties on a series of EuFe$_2$(As$_{1-x}$P$_x$)$_2$ samples and explore the
details of the interplay of AF  and FM  phase of Eu$^{2+}$ moments with  the SC phase as a function of phosphorus doping.
We find that the
SDW transition associated with the Fe moments can be suppressed upon P doping and obtain a bulk SC
phase transition up to 28\,K for $x$=0.2. Further increasing the P content,  SC vanishes
and Eu$^{2+}$ ordering changes from AF to FM. Our results suggest that SC and FM phases compete with each other.
Careful analysis also shows that the bulk SC phase co-exists with Eu
AF phase, possibly in orthorhombic symmetry. 
Density functional theory based calculations also witnesses a change in the ordering of the Eu$^{2+}$ moments
from AF to FM with increasing phosphorus content. Further analysis allows us to infer that the lattice effects is
more conducive to the AF to FM transformation of the Eu$^{2+}$ moments, rather than the phosphorus substitution itself.
More microscopic experiments like $\mu$SR, low temperature
powder diffraction, NMR etc. are currently in progress to confirm the different phases reported here.

The authors would like to thank, C. Geibel, Y. Tokiwa and K. Winzer for discussion and help. We acknowledge
financial support by the DFG Research Unit SPP-1458.


\bibliography{ref.bbl}

\begin{thebibliography}{99}
\bibitem{foot}Note that in Ref.~\onlinecite{ahmed}, despite the coexistence of SC and Eu-FM order,
a competition of SC and Fe-FM was claimed.
\bibitem{footnote}Another way of obtaining an estimate for `U' is the energy difference of the spectral weight transfer due to a valence
change of Eu from a 2+ to a 3+ state. A recent report (Ref.~\onlinecite{sun}) evidences such a valence transition under pressure for 
EuFe$_{2}$As$_{2}$ and consistent with our assumptions for `$U_{\rm 4f}$', the energy difference between the main Eu$^{2+}$ peak and the satellite Eu$^{3+}$ peak is $\approx$ 8 eV.
\end{thebibliography}

\end{document}